\documentclass[3p,sort&compress,fleqn,authoryear]{elsarticle}
\usepackage{amssymb,amsmath,latexsym}
\usepackage[utf8]{inputenc}
\usepackage{pifont}   % For openstar in the title footnotes
\usepackage{geometry} % For margin settings
\usepackage{graphicx} % For graphics inclusions
\usepackage[dvipsnames]{xcolor}
\usepackage{bm}
\usepackage[breaklinks=true,colorlinks=true]{hyperref}
\hypersetup{
	allcolors       = {black},
	linkbordercolor = {white},
	linkcolor       = {brown},
	urlcolor        = {blue},
    citecolor       = {blue}}
\usepackage{epstopdf}
\usepackage{lineno}   % line numbers
\modulolinenumbers[5] 

\newcommand{\defi}{:=}

\newcommand{\pv}{\mathop{\mathrm{pv}}}
\newcommand{\Pf}{\mathop{\mathrm{Pf}}}
\newcommand{\sign}{\mathop{\mathrm{sign}}}
\renewcommand{\Im}{\mathop{\mathrm{Im}}}
\renewcommand{\Re}{\mathop{\mathrm{Re}}}
\newcommand{\dd}{\mathrm{d}}
\newcommand{\ee}{\mathrm{e}}
\newcommand{\ii}{\mathrm{i}}
\newcommand{\bigO}[1]{\mathrm{O}\left(#1\right)}
\newcommand{\ba}{\mathbf{a}}
\newcommand{\bb}{\mathbf{b}}
\newcommand{\bk}{\mathbf{k}}
\newcommand{\bl}{\mathbf{l}}
\newcommand{\bbm}{\mathbf{m}}
\newcommand{\bn}{\mathbf{n}}
\newcommand{\br}{\mathbf{r}}
\newcommand{\bt}{\mathbf{t}}
\newcommand{\bu}{\mathbf{u}}
\newcommand{\bv}{\mathbf{v}}
\newcommand{\bx}{\mathbf{x}}
\newcommand{\bA}{\mathbf{A}}
\newcommand{\bL}{\mathbf{L}}

\newcommand{\bV}{\mathbf{V}}

\newcommand{\bhb}{\mathbf{\widehat{b}}}

\newcommand{\bhe}{\mathbf{\widehat{e}}}
\newcommand{\bhk}{\mathbf{\widehat{k}}}

\newcommand{\sfB}{\mathsf{B}}
\newcommand{\sfC}{\mathsf{C}}

\newcommand{\sfG}{\mathsf{G}}

\newcommand{\sfI}{\mathsf{I}}
\newcommand{\sfK}{\mathsf{K}}
\newcommand{\sfL}{\mathsf{L}}

\newcommand{\sfN}{\mathsf{N}}
\newcommand{\sfP}{\mathsf{P}}
\newcommand{\bbeta}{\bm{\eta}}
\newcommand{\bsigma}{\bm{\sigma}}
\newcommand{\bzeta}{\bm{\zeta}}
\newcommand{\cS}{c_{\rm S}}
\newcommand{\cL}{c_{\rm L}}

\newcommand{\calN}{\mathcal{N}}
\newcommand{\calS}{\mathcal{S}}

\renewcommand\atop[2]{\genfrac{}{}{0pt}{}{#1}{#2}}

\newcommand{\YP}[1]{\begingroup\color{ForestGreen}#1\endgroup}
\begin{document}
\begin{frontmatter}
	\title{Dynamic stress response kernels for dislocations and cracks: unified anisotropic Lagrangian formulation\\ 
	(October 28, 2025)}
		
	\author[dif,lmce]{Yves-Patrick Pellegrini}
	\ead{yves-patrick.pellegrini@cea.fr}
	\author[cad]{Marc Josien}
	\ead{marc.josien@cea.fr}
	\author[enpc]{Martin Chassard}
	%\ead{martin.chassard@cea.fr}
		
	\address[dif]{CEA, DAM, DIF, F-91297 Arpajon Cedex, France.}
	\address[lmce]{Université Paris-Saclay, CEA, LMCE, F-91680 Bruyères-le-Châtel, France.}
	\address[cad]{CEA, DES, IRESNE, DEC, SESC, LMCP, Cadarache, F-13108, Saint-Paul-Lez-Durance, France.}
	\address[enpc]{\'Ecole Nationale des Ponts et Chauss\'ees, 6 et 8 avenue Blaise Pascal, F-77455 Marne-La-Vallée Cedex 2, France.}
		
	\date{October 28, 2025}
		
	\begin{abstract}
	Elastodynamic cohesive-zone models for defects such as cracks or dislocations (such as the Geubelle-Rice model for cracks, or the Dynamic Peierls Equation for flat-core dislocations), feature the same stress-response convolution kernel in space and time. It accounts for in-plane elastic wave propagation, while its associated instantaneous radiative term accounts for radiative losses in the surrounding medium. These objects are well-known for isotropic elasticity, with their space-time representations involving generalized functions. For anisotropic elasticity they were unknown. The paper presents a derivation using the Stroh formalism. Their Fourier representation rests exclusively on the so-called prelogarithmic Lagrangian factor $L(v)$, while their space-time form involves its derivative $p(v)=L'(v)$, the prelogarithmic impulsion function.
    A straightforward consequence is the reformulation of the stress in the Weertman model of steadily-moving dislocations in terms of $L(v)$. Special care being paid to the causality constraint, the theory covers indifferently subsonic, intersonic and supersonic regimes of motion. The theory proposed is suitable to phase-field-type Fourier-based numerical codes for planar systems of defects in anisotropic elastodynamics. 
	\end{abstract}
\end{frontmatter}
	
%%%%%%%%%%%%%%%%%%%%%%%%%%%%%%%%%%%%%%%%%%%%%%%%%%%%%%%%%%%%%%%%%%%%%%%%%%%%%%%%%%%%%%%%%%%
\section{Introduction}
Since atomistic simulations \citep{GUMB99b,JING08,OLMS05,TSUZ08,DAPH14} and experimental measurements \citep{ROSA99,NOSE07,FARA10,JARA21,KATA23} have confirmed the possibility of intersonic\footnote{
Here, we use the term intersonic \citep{SAMU02} to refer to velocities between the smallest and largest wave speeds; the term transonic is avoided, as it has a specific meaning in shock-wave physics, and refers to specific isolated velocity states in the context of surface-wave propagation \citep{TING96}.} or supersonic motion, elastodynamic radiation properties of fast moving line defects such as dislocations or cracks \citep{ZHAN15} have attracted renewed attention \citep{GURR16,PELL18,BLAS23b}, often in connection with their steady-state mobility law \citep{ROSA01,TADU23,BLAS23a}, or their equation of motion \citep{ESHE53,PILL07,PELL14,GURR16}. Yet, many theoretical studies are still carried out in isotropic elasticity while realistic applications require considering elastic anisotropy \citep{BULL54,TEUT63b,WEER62a,BACO80}, the dynamics of which can be handled efficiently through the causal version \citep{PELL17a} of the Stroh formalism \citep{STRO62,TING96,BARN73a,BARN02,TANU07}.
 
The so-called \emph{prelogarithmic Lagrangian factor} function $L(v)$ \citep{BELT68}, where $v$ is the defect velocity, has long been known to play a key role in the elastodynamics of moving defects \citep{MALE70c,BARN73b,HIRT98,WU02}. In an approximate equation of motion for dislocations in isotropic elasticity, an analytic continuation of $L(v)$ to the upper complex plane of dislocation velocities was introduced \citep{PELL12,PELL14}. Still, the significance of imaginary part of $L(v+\ii 0^+)$, while related to causality and radiative dissipation, is not fully understood. Also, the local instantaneous radiative term in the dynamical Peierls model \citep{PELL23} and in fracture models \citep{COCH94,GEUB95} needs further clarification, as its anisotropic expression is still unknown. We confine ourselves to continuum mechanics, leaving  outside the scope of the paper temperature \citep{GURR17a} and lattice-dispersion effects.

Focusing on the in-plane stress, the present work fully elucidates the above issues, while exploring new avenues. Specifically, we demonstrate, within the framework of anisotropic elasticity, that the entire theory of the radiative in-plane stress response of flat moving defects---dislocations or cracks---can be formulated \emph{exclusively} in terms of the function $L(v)$ and its first derivative (the impulsion function), or through its associated (Lagrangian) energy tensor. All of these quantities are obtained from the Stroh formalism. The technique differs from previous pioneering works on radiation by dislocations in anisotropic media \citep{MARK87,PAYT85b}. Remarkably, owing to a well-known homogeneity property \citep{WU00b}, these functions appear both in the real space–time domain ($x,t$) and in the Fourier domain of wavevector–frequency ($\bk,\omega$), with $v$ being replaced by $x/t$ or $\omega/k$, respectively. In the Fourier representation \citep{MURA87}, we employ the Stroh formalism to derive the radiative stress response of an ensemble of dislocations or cracks distributed in the plane. 

Our formulation covers any speeds, including supersonic of intersonic ranges, and is targeted to applications to cohesive-zone-type models \citep{GEUB95,SAMU02,ZHOU05}, especially with regard to phase-field type numerical methods of solutions based on fast Fourier transforms (FT) in the plane, and/or Laplace transforms in time \citep{ROCH22,PELL23}. For dislocations the appropriate context is that of generalized Peierls-Nabarro-type models \citep{PEIE40,NABA47}, which are cohesive-zone models in shear \citep{DENO04,HUNT11,BEYE16}. For a related elastodynamic models for dislocations see, e.g., \cite{ACHA25}.

The paper is organized as follows. Section \ref{sec:traction} revisits the classical calculation of the tensor kernel for vector traction generated by a slip on the slip plane, using one Stroh-type eigensolution for each in-plane Fourier wave vector while accounting for causality. The Section closes by establishing symmetry properties under frequency and wavector inversion, and by the computation of the anisotropic radiative coefficient---shown to originate from a limiting behavior when $|\omega/k|\to +\infty$. 

Section \ref{sec:straightdef} particularizes the results to straight defects. Uniform steady motion is considered in Sec.\ \ref{sec:steady} where the stress kernel of the Weertman equation \citep{WEER69a,ROSA01,JOSI18a,JOSI19} is obtained in terms of $L(v)$ for the first time. The time-dependent dynamical problem is considered next, and the dynamical response in space-time form is established in terms of the function $p(v)=L'(v)$. In these derivations, the prelogarithmic Lagrangian factors shows up in the stress response function, endowed with an imaginary part, although its primary definition is energetic. 

In Section \ref{sec:plfr} investigates this further and demonstrates, starting from the Green's function of the Navier equation in the Fourier representation, that the causal stress-response kernel differs from the kernel in the Lagrangian only by its radiative part. This explains why retrieving the stress response from the Lagrangian kernel requires re-instating causal radiative properties by means of an analytical continuation in the upper complex plane of velocities. We conclude in Sec.\ \ref{sec:concl}.  

\ref{sec:UFT} states our Fourier-transform (FT) conventions and recalls useful formulas.

%%%%%%%%%%%%%%%%%%%%%%%%%%%%%%%%%%%%%%%%%%%%%%%%%%%%%%%%%%%%%%%%%%%%%%%%%%%%%%%%%%%%%%%%%%%
\section{Traction and resolved-stress kernel from Stroh formalism}
\label{sec:traction}
%******************************************************************************************
\subsection{Outline}
The model consists of two elastically anisotropic half-spaces of same material with elastic tensor $c_{ijkl}=c_{jikl}=c_{klij}$ and material density $\rho$, separated by a plane interface at $z=0$  of normal $\bn$ along the $Oz$ coordinate axis. The plane (referred as such hereafter) is the crack plane, or the glide plane of the dislocation. It is spanned by physical-space coordinates $\br=(x,y)$, and $t$ is the time. The stress is $\sigma_{ij}=c_{ijkl}\partial_k u_l$, and the traction vector $\bt$ on the plane $z=0$ has components $t_i=\sigma_{ij}n_j$. Denoting by $\bu$ the material-displacement vector, time-dependent boundary conditions are standard ones \citep{MURA63a,NAKA73,WU02} of prescribed displacement discontinuity $\bbeta$ across the plane, which represents the local plastic slip or the crack opening, and of continuity of tractions,  namely,
\begin{subequations}
\begin{align}
\label{eq:displcond}
\bbeta(\br,t)&=\lim_{h\to 0^+}\left[\bu(\br,z+h,t)-\bu(\br,z-h,t)\right],\\
\label{eq:tractcond}
\bt(\br,z=0^+,t)&=\bt(\br,z=0^-,t)\equiv\bt(\br,t).
\end{align}
\end{subequations}
More restrictive conditions are sometimes used; e.g., \cite{BLAS23b}. 
In each half-space, $\bu$ obeys the Navier equation
\begin{align}
\label{eq:navier}
c_{ijkl}\partial_j\partial_k u_l&=\rho\,\partial_t^2 u_{i}.
\end{align}
For cracks, the total traction --which includes the additive contribution of the externally-applied stress-- should vanish on the crack faces. In a cohesive-zone approach---and unlike in classical calculations---the latter feature is not imposed as a boundary condition in \eqref{eq:tractcond}, but rather emerges from the opening-dependent stress balancing term in the model, which vanishes beyond a maximal opening \citep{GEUB95,SAMU02}. For dislocations the balancing term is the pull-back force that derives from the generalized stacking fault ($\gamma-$surface) potential \citep{MRYA98,ALBR16} in a generalized Peierls approach. This component of the model, necessary to complete the equation for $\bbeta$, relates to nonlinear material response in the cohesive zone and lies out of the scope of the present work. The above formulation applies to planar system of cracks or dislocations \citep{MURA63a}. For dislocations $\bbeta$ can be a three-dimensional (3D) vector \citep{HUNT11}.

The linear elastodynamic kernel that relates $\bt$ to $\bbeta$ is derived below in Fourier form using the Stroh formalism. Initially aimed at solving plane-strain problems of elastic fields from straight sources, the Stroh formalism has been combined with the Radon transform into a 3D formalism \citep{WU98}---including extensions to cylindrical and polar coordinates under specific conditions \citep{NORR12}, and has found numerous applications in various areas of engineering mechanics \citep{HWUB22}. Our calculation parallels the usual one for a steadily-moving singularity at speed $v$, with important differences: (i) it applies to a system of non-necessarily straight defects, the calculation being generalized to a full two-dimensional in-plane (2D) setting by means of FTs; (ii) time-dependent aspects will be addressed by replacing the steady-state velocity $v$ by $\omega/k+\ii 0^+$, where $k>0$ is the in-plane wave vector modulus, $\omega$ is the angular frequency, and the vanishingly small imaginary part implements causality. 
	
%******************************************************************************************
\subsection{Two-dimensional in-plane derivation}
\label{sec:2dinplane}
In the plane $z=0$ we use a space-time FT of two-dimensional wave vector $\bk=(k_x,k_y)=k\,\bhk$, where $\bhk$ is the director, and $\smash{k=(k_x^2+k_y^2)^{1/2}}$. No FT is taken along the $z$-direction. Following the notation of \cite{BARN73b}
\begin{align}
\label{eq:notabraces}
(\ba\bb)_{ij}&\defi a_k c_{iklj} b_l    
\end{align}
for any two vectors $\ba$ and $\bb$,  denoting the identity matrix by $\sfI$, letting $k_x=k_1$ and $k_y=k_2$ and assuming  that $k\not=0$, Eq.\ \eqref{eq:navier} becomes
\begin{align}
\label{eq:navierkom}
\left\{(\bhk\bhk)+[(\bhk\bn)+(\bn\bhk)]\left(\frac{\partial_z}{\ii k}\right)+(\bn\bn)\left(\frac{\partial_z}{\ii k}\right)^2-\rho\left(\frac{\omega}{k}+\ii 0^+\right)^2\,\sf{I} \right\}\cdot\bu(\bk,\omega,z)=0,
\end{align}
which depends on $\omega$ only via the ratio $\omega/k$. The limiting case 
$k=0$ is addressed in Section \ref{sec:trltc}. 

Eq.\ \eqref{eq:navierkom} implements the principle of limiting absorption (PLA) \cite[pp.\ 62--65]{HARR01}, whereby $\omega$ is replaced by $\omega+\ii 0^+$, with $0^+$ a vanishingly small positive number (of irrelevant exact magnitude), as usual in dynamics ; e.g., \cite{RAMA97}. Thus $\omega/k$ becomes $\omega/k+\ii 0^+$. The response kernel that stems from \eqref{eq:navierkom} is thus analytic in the upper complex half-place  $\Im\omega\geq 0$ with real axis included, which ensures causality by the Paley-Wiener or Tichtmarsch theorems 
\citetext{\citealp[p.~407]{BART89}; \citealp{BUCH59}; \citealp{TOLL56}}. The ensuing wave solutions satisfy the Sommerfeld outgoing--radiation condition \citep{RODN15}. The PLA automatically allows proper rendering of radiative drag via Mach-cones in intersonic or supersonic steady-state regimes \citep{LAZA16,PELL17a}.

To solve \eqref{eq:navierkom} the Stroh method \citep{STRO58,STRO62} introduces six eigenvalues $p_\alpha$ and eigenvectors $\bA^\alpha$, $\alpha=1,\ldots,6$, such that
\begin{align}
\label{eq:stroheq}
\left\{(\bhk\bhk)+[(\bhk\bn)+(\bn\bhk)]p_\alpha+(\bn\bn)p_\alpha^2-\rho\left(\frac{\omega}{k}+\ii 0^+\right)^2\,\sf{I}\right\}\cdot\bA^\alpha=0.
\end{align}

Introducing (formally, for now) the shorthand notation $v=\omega/k$, the $p_\alpha$s are determined by the sextic equation
\begin{align}
	\label{eq:sextic}
	\det\left\{(\bhk\bhk)+[(\bhk\bn)+(\bn\bhk)]p_\alpha+(\bn\bn)p_\alpha^2-\rho\,(v+\ii 0^+)^2\sf{I}\right\}=0,
\end{align}

Equation \eqref{eq:sextic} resembles the one for a straight dislocation of velocity $\bv = v\,\bbm$ in direction, but with $\bbm$ replaced by $\bhk$. If we denote by $\widetilde{\bk}=(k_x,k_y,k_z)$ the full three-dimensional wave vector, with given real-valued $k_x$ and $k_y$, and $k:=(k_x^2+k_y^2)^{1/2}$, finding the $p_\alpha$ in Eq.\ \eqref{eq:stroheq} amounts to finding complex-valued solutions $k_z$ of the dispersion relation
\begin{align}
\det\left[\widetilde{k}_j c_{ijkl}\widetilde{k}_k-\rho\,(\omega+\ii 0^+)^2\sf{I}\right]=0,
\end{align}
letting $k_z\equiv p_\alpha k$.
%\MJ{under the constraints $k_x \in \mathbb{R}$, $k_y \in \mathbb{R}$, $k_z = p_\alpha \sqrt{k_x^2 + k_y^2}$.}
%\footnote{Using 3D polar coordinates where $\widetilde{k}=\widetilde{k}(\sin\theta\cos\phi,\sin\theta\sin\phi,\cos\theta)$ and the definition of $k$ one deduces that $p=\cot\theta$.} 
These solutions define plane-wave modes $\exp(\ii\,k\,p_\alpha z)$. If $\Im p_\alpha\not=0$ they are localized near to the surface, representing the field of a moving subsonic defect. They turn into bulk modes in inter/supersonic ranges of $v$ where pairs of the $p_\alpha$ become real (in absence of $\ii 0^+$). Mach cones that extend to infinity are made of special bulk modes endowed with a causal character. Causality selects branches with correct orientation with regard to the direction of motion.

Stroh transforms the problem of finding the pairs $(p_\alpha,\bA^\alpha)$ into one of 6-dimensional spectral decomposition. In its modern form \citep{BARN73b,MALE71a,NAKA97,LOTH09} the technique amounts to solving the eigenvalue problem $\calN\cdot\bzeta^\alpha=p_\alpha\bzeta^\alpha$, where $\calN$ is the $6\times 6$ matrix
\begin{align}\label{def:calN}
\calN(v)&:=
\begin{pmatrix}
-(\bn\bn)^{-1}\cdot(\bn\bhk) & -(\bn\bn)^{-1} \\
-(\bhk\bn)\cdot(\bn\bn)^{-1}\cdot(\bn\bhk)+(\bhk\bhk)-\rho v^2\sf{I} & -(\bhk\bn)\cdot(\bn\bn)^{-1}
\end{pmatrix}.
\end{align}
The last three components of the 6-eigenvectors $\bzeta^\alpha \equiv(\bA^\alpha,\bL^\alpha)$ define traction vectors $\bL^\alpha$, such that
\begin{align}
\label{eq:tract}	
\bL^\alpha=-[(\bn\bhk)+p_\alpha (\bn\bn))]\cdot\bA^\alpha.
\end{align}
By \eqref{eq:stroheq} and \eqref{eq:tract} the eigenvalues $p_\alpha$, $\bA^\alpha$, and $\bL^\alpha$ exclusively depend on the ratio $\omega/k$, and on $\smash{\bhk}$; the latter dependence will be left implicit hereafter. It is in general possible to enforce the normalization \begin{align}
\label{eq:normaliz}	
\bA^\alpha\cdot\bL^\beta + \bA^\beta\cdot\bL^\alpha=\delta_{\alpha\beta},    
\end{align}
save for isolated values of $v$ for which $\bA^\beta\cdot\bL^\alpha=0$ for some $\alpha$. Problematic cases also arise when matrix $\mathcal{N}$ is non-semisimple. Such degenerate cases correspond to isolated velocity states that have been extensively studied \citep{CHAD77b,TANU07,LOTH09}.
%Unlikely to be met in a numerical method of solution, they are not further considered.
Hence, we do not further consider these degeneracies.
For use below, we mention another important identity \citep{NISH72}, which in the present context reads
\begin{align}
\label{eq:pLLident}	
&-[(\bhk\bn)\cdot(\bn\bn)^{-1}\cdot(\bn\bhk)-(\bhk\bhk)
+\rho\,v^2\,\mathsf{I}]=\sum_{\alpha=1}^6 p_\alpha\bL^\alpha\otimes\bL^\alpha.
\end{align}
	 
Due to the PLA, the problem solved in this paper reads $\calN(v+\ii 0^+)\cdot\bzeta^\alpha=p_\alpha\bzeta^\alpha$, and eigenvalues $p_\alpha$ always have at least an infinitesimal (nonzero) imaginary part, which is a notable difference from the classical approach. More precisely, the perturbation $\delta v\defi\ii \epsilon$ with $\epsilon\to 0^+$ imparts an imaginary correction $\delta p_\alpha \propto\epsilon$ to the real inter-/supersonic modes (\ref{app:pert}). To build localized field solutions, we introduce the signs $s_\alpha\defi\sign\Im p_\alpha$, and categorize eigenvalues and eigenvectors so that $s_\alpha=+1$ for $\alpha\in \calS^+$ and $s_\alpha=-1$ for $\alpha\in\calS^-$, where $\calS^\pm$ are disjoint subsets of indices such that  $\calS^+\cup\calS^-=\{1,2,3,4,5,6\}$. 
%%%%%%%The following derivation does not need that $\calS^+$ and $\calS^-$ each contain exactly 3 elements.
%\footnote{This should hold anyway, since the vectors $\bA^\alpha$ should, as eigenvectors, define a full polarization basis for the displacement vector in each half-space $z\gtrless 0$.}
%\footnote{\MJ{Phrase supprimée : Je n'insisterai pas là-dessus.}}

Equation \eqref{eq:navier} admits exponentially-decaying (i.e., surface-wave) solutions localized around the glide plane $z=0$ of the type \citep{NAKA73}
\begin{align}
\label{eq:displ}
\bu(\bk,z,\omega)=
\begin{cases}
	\sum_{\alpha\in\calS^+} a_\alpha \bA^\alpha \ee^{\ii k p_\alpha z}\qquad\text{if}\qquad z>0 \\
		\sum_{\alpha\in\calS^-} a_\alpha \bA^\alpha \ee^{\ii k p_\alpha z}\qquad\text{if}\qquad z<0,
\end{cases}
\end{align}
where coefficients $a_\alpha$ are to be determined. Given $z$, the subset of coefficients with indices $\alpha\in\calS^+$ on the one hand, and that with indices $\alpha\in\calS^-$ on the other hand, are never at play simultaneously. Thus, we avoid introducing distinct symbols for the unknowns $a_\alpha$ in each of the half-spaces.

With \eqref{eq:displ}, the Fourier form of the displacement discontinuity \eqref{eq:displcond} reads
\begin{align}
\label{eq:slip}
\bbeta(\bk,\omega)&:=\bu(\bk,0^+,\omega)-\bu(\bk,0^-,\omega)=\sum_{\alpha=1}^6 s_\alpha a_\alpha  \bA^\alpha.
\end{align}
From \eqref{eq:tract} and \eqref{eq:displ} the tractions $t_i(\bk,z,\omega)=n_j\sigma_{ij}=n_j c_{ijkl}\partial_k u_l(\bk,z,\omega)$ along the slip-plane normal read
\begin{align}
\label{eq:tractions}
\bt(\bk,z,\omega)
&=-\ii k
\begin{cases}
	\sum_{\alpha\in\calS^+} a_\alpha \bL^\alpha\,\ee^{\ii k p_\alpha z}\qquad\text{if} \qquad z>0,\\
	\sum_{\alpha\in\calS^-} a_\alpha \bL^\alpha\,\ee^{\ii k p_\alpha z}\qquad\text{if} \qquad z<0.
\end{cases}	
\end{align}
Meanwhile, the continuity condition \eqref{eq:tractcond} reads $\sum_{\alpha\in\calS^+} a_\alpha \bL^\alpha=\sum_{\alpha\in\calS^-} a_\alpha \bL^\alpha$, or
\begin{align}
	\label{eq:ident1}
	\sum_{\alpha=1}^6 s_\alpha a_\alpha \bL^\alpha=0.
\end{align}	
Consequently, the traction on the slip plane $z=0$ is 
\begin{align}
	\label{eq:traction_plane}	
	\bt(\bk,\omega)&=-\frac{\ii k }{2}\sum_{\alpha=1}^6 a_\alpha \bL^\alpha.
\end{align}	
Coefficients $a_\alpha$ follow from taking the dot product of \eqref{eq:slip} with $\bL^\beta$, transforming the result using \eqref{eq:normaliz}, and simplifying thanks to \eqref{eq:ident1}, to give
\begin{align}
	\bL^\beta\cdot\bbeta
	&=\sum_{\alpha=1}^6s_\alpha a_\alpha\bA^\alpha.\bL^\beta=s_\beta a_\beta-\bA^\beta\cdot\sum_{\alpha=1}^6s_\alpha  a_\alpha \bL^\alpha=s_\beta a_\beta,	
\end{align}	
whereby
\begin{align}
\label{eq:alphaval}
a_\alpha = s_\alpha \bL^\alpha\cdot\bbeta.	
\end{align}
Substituting \eqref{eq:alphaval} into \eqref{eq:traction_plane} yields the desired 2D traction-discontinuity relationship as
\begin{subequations}
\begin{align}
\label{eq:tractvseta2}	
\bt(\bk,\omega)
&=(2\pi) k\,\sf{L}\left(\bhk,\frac{\omega}{k}+\ii 0^+\right)\cdot\bbeta(\bk,\omega),\hspace{4cm}(2D)\\
\label{eq:Ldef}	
\sfL(\bhk,v)
&:=\frac{1}{4\ii\pi}\sum_{\alpha=1}^6 s_\alpha\bL^\alpha\otimes\bL^\alpha.	
\end{align}
\end{subequations}
It this expression, we have introduced for convenience the matrix operator $\sf{L}$, related to the conventional matrix $\sfB$ \citep{LOTH92c,LOTH92d} by $\sfL=-\sfB/(4\pi)$. The operator $\sfL(\bhk,v)$ is an analytic function of $v$ in the upper half-complex plane. Its principal determination exhibits branch cuts along the real axis, located in the overlapping regions 
$v\in]-\infty,-c_\gamma(\bhk)]\cup[c_\gamma(\bhk),+\infty[$ where the $c_\gamma(\bhk)$, for $\gamma=1,\ldots, 3$, represent the wave speeds in direction $\bhk$.
With \eqref{eq:Ldef} both writings $\sfL(\bhk,v)$ or $\sfL(\bhk,v+\ii 0^+)$ are equivalent since the infinitesimal $\ii 0^+$ is present from the outset in the formalism. We shall sometimes use the second one to emphasize it. 

The above derivation provides a full dynamical generalization of a previously known result, namely the expression \eqref{eq:Ldef}	of the Lagrangian energy tensor, but in the context of a calculation with FTs. A numerical FT implementation 
%such as \YP{by} \cite{PELL23}, 
would require solving numerically one Stroh eigenproblem per frequency and per in-plane Fourier mode. Of course, with \eqref{eq:alphaval}, expression \eqref{eq:displ} gives access to the out-of plane displacement, deformation, and stress generated from $\bbeta$, if needed. 

%******************************************************************************************
%\subsection{The bicrystal}
%Following \citet{BRAE71} the above derivation admits a straightforward and immediate extension to the problem of an interface dislocation between two anisotropic crystals (1) and (2) of same density $\rho$, where $\mathsf{L}$ in \eqref{eq:tractvseta2} is replaced by
%\begin{align}
%\sf{L}^{1-2}&:=\sf{L}^{(1)}\cdot(\sf{L}^{\rm av})^{-1}
%\cdot\sf{L}^{(2)},
%\end{align}
%where $\sf{L}^{(1,2)}$ for each medium are of the type \eqref{eq:Ldef} and
%\begin{align}
%\sf{L}^{\rm av}&:=\frac{1}{2}\left[\sf{L}^{(1)}+\sf{L}^{(2)}\right].
%\end{align}

%******************************************************************************************
\subsection{Symmetry properties}
Operator $\sfL$ enjoys two key symmetry properties. First, denoting complex conjugation by an overline,
\begin{subequations}
\begin{align}
\label{eq:symmetryv}
\sfL\left(\bhk,-v\right)&=\overline{\sfL}\left(\bhk,v\right).
\end{align}
This is immediate, as $\sfL$ depends on $v$ only through the group $(v+\ii 0^+)^2$. As a result, the real and imaginary parts of  $\sfL(\bhk,v+\ii 0^+)$ are even and odd functions of $v$, respectively. 

Second, 
\begin{align}
\label{eq:symmetry}
\sfL\left(\bhk,v\right)&=\sfL\left(-\bhk,v\right).
\end{align}
\end{subequations}
The proof is straightforward. By \eqref{eq:stroheq}, if $\bhk\to-\bhk$, then $p_\alpha\to-p_\alpha$, and consequently $s_\alpha\to -s_\alpha$. Meanwhile, $\bA^\alpha$ must preserve its direction, so $\bA^\alpha\to \phi_\alpha\bA^\alpha$, where the scalar $\phi_\alpha$ is unknown. Then, by \eqref{eq:tract} we have $\bL^\alpha\to -\phi_\alpha\bL^\alpha$; and from \eqref{eq:normaliz}, it follows that $\phi_\alpha^2=-1$. Combining these transformations within \eqref{eq:Ldef} establishes property \eqref{eq:symmetry} $\square$.

%******************************************************************************************
\subsection{The radiative-loss tensor coefficient}
\label{sec:trltc}
This section addresses the issue of the limit $k\to 0$ in \eqref{eq:tractvseta2}, demonstrating that it defines a pure time-derivative component in kernel $k\,\sfL$. We begin by introducing the constant 
\begin{align}
\sfK:=-2\ii\pi\lim_{k\to 0}\frac{k}{\omega}
\sfL\left(\bhk,\frac{\omega}{k}+\ii 0^+\right),
\end{align}
which we aim to compute. By definition, and with regard to $\sfK$, the limit
\begin{align}
\label{eq:limitkzer}
\lim_{k\to 0} (2\pi) k\,\sfL\left(\bhk,\frac{\omega}{k}+\ii 0^+\right)=\ii\,\omega\, \sfK.
\end{align} 
is equivalent to taking the limit $|\omega|\to\infty$, with $\omega$ in the upper half complex plane. As will be shown, this limit is independent of $\bhk$. To proceed, we examine the leading-order behavior in Eq.\ \eqref{eq:sextic} as $|p_\alpha|\to\infty$:
\begin{align}
\label{eq:defKequ1}	
\det\left[(\bn\bn)p^2-\rho\left(\omega/k+\ii 0^+\right)^2\sfI\right]&=\rho^3\prod_{\gamma=1}^3\left[c_\gamma^{\star\,2} p^2-(\omega/k+\ii 0^+)^2\right]=0,
\end{align}
where the real symmetric matrix $(\bn\bn)$ is diagonalized as 
\begin{align}
(\bn\bn)&=\rho\sum_{\gamma=1}^3c^{\star\,2}_\gamma \bV^\gamma\otimes\bV^\gamma.
\end{align}

The real constants $c_\gamma^\star>0$ represent wave speeds along the direction of the surface normal, and the vectors $\bV^\gamma$ are the corresponding orthonormal eigenvectors. The eigenvalues $\mu_\gamma\defi\rho\, {c_\gamma^\star}^2$ serve as effective elastic moduli.
\iffalse
In the limit, the leading-order equation for $p_\alpha$ becomes
\begin{align}
\label{eq:defKequ1}	
\det\left[(\bn\bn)p^2-\rho\left(\omega/k+\ii 0^+\right)^2\sfI\right]&=\rho^3\YP{\prod_{\gamma=1}^3}\left[c_\gamma^{\star\,2} p^2-(\omega/k+\ii 0^+)^2\right]=0,
\end{align}
\fi
Equation \eqref{eq:defKequ1} has solutions
\begin{align}
\label{eq:psols}	
p_{\gamma} &\defi\frac{\omega}{c_\gamma^\star k}+\ii 0^+,\qquad p_{\gamma+3}\defi-p_{\gamma},\qquad \gamma=1,2,3.
\end{align}
Thus, $s_{\alpha=1,2,3}:=+1$ and $s_{\alpha=4,5,6}:=-1$.
Meanwhile, from  Eq.\ \eqref{eq:stroheq}, $\bA^\alpha$ with $\alpha=1,\ldots,6$ must satisfy
\begin{align}
\label{eq:defKequ2}	
\rho\sum_{\gamma=1}^3\left[c^{\star\,2}_\gamma p_\alpha^2-(\omega/k)^2\right] \bV^\gamma\otimes\bV^\gamma\cdot\bA^\alpha
=\rho(\omega/k)^2\sum_{\gamma=1}^3\left(c_\gamma^{\star\,2}/c_\alpha^{\star\,2}-1\right)\bV^\gamma\otimes\bV^\gamma\cdot\bA^\alpha=0.
\end{align}
Clearly, this equation holds if and only if $\bA^\alpha$ is proportional to $\bV^\gamma$ for $\alpha=\gamma$, and for $\alpha=\gamma+3$ if we extend the indexing of normal wave velocities by letting $c^\star_{\gamma+3}:=c^\star_\gamma$. The proportionality factor is determined up to an irrelevant sign by the normalization condition \eqref{eq:normaliz}, which yields, under same conditions on indices, $\bA^\alpha=\ii\,\bV^\gamma/\sqrt{2 p_\alpha\mu_\gamma}$ and $\bL^\alpha=-p_\alpha\mu_\gamma\bA^\alpha=-\ii\,\sqrt{p_\alpha\mu_\gamma/2}\,\bV^\gamma$. Consequently, in the limit,
\begin{align}
-2\ii\pi\lim_{k\to 0}\frac{k}{\omega}\frac{1}{4\ii\pi}\sum_{\alpha=1}^6 s_\alpha \bL^\alpha\otimes\bL^\alpha
&=\lim_{k\to 0}\frac{k}{\omega}\frac{1}{4}\sum_{\gamma=1}^3 \left(s_\gamma p_\gamma+s_{\gamma+3} p_{\gamma+3}\right)\mu_\gamma \bV^\gamma\otimes\bV^\gamma.	
\end{align}
Upon substituting the expressions for $\mu_\gamma$ and $p_\alpha$ given above, this simplifies to the positive-definite tensor
\begin{align}
\label{eq:sfKdef}
\sfK&=\frac{\rho}{2}\sum_{\gamma=1}^3c^\star_\gamma\,\bV^\gamma\otimes\bV^\gamma =\frac{\sqrt{\rho}}{2}(\bn\bn)^{1/2},
\end{align}
which does not depend on $\bhk$, as announced.
The infinitesimals in \eqref{eq:psols} have been dropped here as no singularity is involved. 
%as this quantity is absent from the defining equations \eqref{eq:defKequ1} and \eqref{eq:defKequ2}.

In the space-time representation, with $\bx=(x,y)$, the term $\ii\omega \,\sfK\cdot\bm{\eta}(\bk,\omega)$ corresponds to a local radiative-drag traction \citep{COCH94,PELL10} of the form
\begin{align}
\label{eq:dragterm}
\bt^{\rm rad}(\bx,t):=-\sfK\cdot\frac{\partial\bm{\eta}}{\partial t}(\bx,t).
\end{align}
For isotropic elasticity, where the elastic stiffness tensor is given by
\begin{align}
c_{ijkl}=\lambda\delta_{ij}\delta_{kl}+\mu(\delta_{ik}\delta_{jl}+\delta_{il}\delta_{jk}),
\end{align}
with $\mu$ and $\lambda$ the Lam\'e moduli, and wave speeds $\cS=\sqrt{\mu/\rho}$ (shear) and $\cL=\sqrt{(\lambda+2\mu)/\rho}$ (longitudinal), a straightforward calculation yields
\begin{align}
\sfK&=\sfK^{\rm iso}:=\frac{\mu}{2\cS}\left[\frac{\cL}{\cS}\bn\otimes\bn+(\sfI-\bn\otimes\bn)\right].	
\end{align}
The tensor components within the brackets correspond precisely to the isotropic dimensionless $\kappa$-coefficients associated with radiative terms for the three dislocation characters, as defined in Eqs.\ (A4) of \citep{PELL23}: $\kappa=1$ for screw and gliding edge dislocations, and $\kappa=\cL/\cS$ for climbing edges.

%\YP{[Paragraph with definition of $\sfC$ removed, as it is not used as such below.]}
%By subtracting this drag term, we define our final in-plane response kernel as
%\begin{align}
%\label{kernelC}
%\mathsf{C}(\bhk,\MJ{\omega/k})&:=(2\pi)\,k\,\sfL(\bhk,\MJ{\omega/k}+\ii\, 0^+)-\ii\,\omega\,\sfK,
%\end{align}
%By construction this kernel vanishes as $|\omega|\to\infty$ in the upper half of the complex $\omega$-plane \MJ{(see \ref{sec:limitLcorr})}. This asymptotic behavior is essential for deriving the space-time representation of the stress kernel; see Section \ref{sec:dsrstf} below.

%%%%%%%%%%%%%%%%%%%%%%%%%%%%%%%%%%%%%%%%%%%%%%%%%%%%%%%%%%%%%%%%%%%%%%%%%%%%%%%%%%%%%%%%%%%
\section{Straight defects (1D case): steady state and dynamics, in isotropic-like form}
\label{sec:straightdef}
In the framework of the Peierls–Nabarro model, the steady state of uniform motion of a straight dislocation is governed by the Weertman equation \citep{WEER69a, ROSA01, JOSI18a, JOSI19, PELL23}. This model provides a foundation for a physically-consistent stress–velocity (mobility) law for ultra-fast dislocations \citep{ROSA01}. This section examines the steady-state and dynamic kernels in space-time form, in terms of the prelogarithmic Lagrangian factor and its derivative.

%******************************************************************************************
\subsection{One-dimensional dynamic response in Fourier form, and prelogarithmic Lagrangian factor}
The one-dimensional (1D) dynamical kernel for a straight, planar, dislocation or crack can be directly derived from the 2D formulation. In the 1D setting, we have $k_y=0$, so $\bk = \bbm\, k_x$, $k=|k_x|$ and $\bhk = \sign(k_x) \bbm$, where the unit vector $\bbm$ is along the $Ox$ axis. Owing to symmetry property \eqref{eq:symmetry} the sign of $k_x$ can be ignored. Eliminating the now irrelevant dependence on $\bhk$ and substituting the expression for $k$ into  \eqref{eq:tractvseta2} yields
\begin{align}
\label{eq:tractvseta1d}	
\bt(k_x,\omega)
&=(2\pi) |k_x|\,\sfL\left(\frac{\omega}{|k_x|}+\ii 0^+\right)\cdot\bbeta(k_x,\omega),\quad\text{with}\quad
\sfL(v)
:=\frac{1}{4\ii\pi}\sum_{\alpha=1}^6 s_\alpha\bL^\alpha\otimes\bL^\alpha.\qquad\text{(1D)},
\end{align}
where the rightmost expression defines the 1D kernel. 

For dislocations, the main issue resides in computing the resolved stress, $\sigma_{\rm r}:=\bt\cdot\bhb$ on the slip system $(\bn,\bhb)$, where $\bhb$ is the direction of the in-plane Burgers vector $\bb$ of (ideal) perfect dislocations. Furthermore, in the simplest version of the Peierls model, $\bbeta=\bhb\,\eta$, where $\eta$ is a scalar. 
%One gets from \eqref{eq:tractvseta2} 
%\begin{align}
%	\sigma_{\rm r}(\bk,\omega)&=(2\pi)k\,\bhb\cdot\sfL\left(\bhk,\frac{\omega}{k}+\ii 0^+\right)\cdot\bbeta(\bk,\omega).
%\end{align}.
The resolved stress can then be conveniently written in terms of a scalar response kernel
\begin{align}
\label{eq:sigr_scal}
\sigma_{\rm r}(k_x,\omega)&=\frac{2\pi}{b^2}|k_x|\,L\left(\frac{\omega}{|k_x|}+\ii 0^+\right)
\eta(k_x,\omega),
\end{align}
where we have introduced the \emph{prelogarithmic Lagrangian factor} function \citep{MALE70c},
\begin{align}
\label{eq:scalLdef}	
L(v)&:=\bb\cdot\sfL(v)\cdot\bb=\bb\cdot\left(\frac{1}{4\ii\pi}\sum_{\alpha=1}^6 s_\alpha\bL^\alpha\otimes\bL^\alpha\right)\cdot\bb
\end{align}
This function enters the expression of the Lagrangian density $\mathcal{L}(v)$ of a straight Volterra dislocation moving with steady velocity $v$. By definition, $\mathcal{L}(v)$ is the difference between the kinetic and elastic energy densities, per unit dislocation length, and evaluates to \citep{STRO62,BARN73a}
\begin{align}
\label{eq:prelogdef}
\mathcal{L}(v)&=L(v)\log(R/r_0),     
\end{align}
where the outer and inner cut-off radii $R$ and $r_0$ with $R/r_0\gg 1$, delimit the tubular domain in which linear elasticity holds, and in which the energy is evaluated. The function $\mathcal{L}(v)$ is further considered in Section \ref{sec:lagsection} below, starting from the Green's function representation of elastodynamic fields \citep{MURA63a}.
%and reads
%\begin{align}
%\label{eq:calLdef}
%\mathcal{L}(v)=L(v)\log(R/r_0)=\int_{r_0^2\leq(x-vt)^2+z^2\leq R^2}\dd^2\bx\,\left(\frac{\rho}{2}\dot{u}_i^2-\frac{1}{2} u_{i,j}c_{ijkl} u_{k,l}\right),
%\end{align}
%where $\bx=(x,z)$ is in the sagittal plane (transverse to the defect), $\bu(x,z,t)=\bu(x-vt,z)$, and $R$ and $r_0$ are outer and inner cut-off radii such that $R/r_0\gg 1$. They delimit the tubular domain in which linear elasticity holds and tame the logarithmic divergence of the integral. 

The present $L(v)$ and $\mathcal{L}(v)$ have been sometimes denoted $L(v)$ and $\Lambda(v)$ \citep{BELT68}, while $p(v)$ usually corresponds to the impulsion---the first derivative of $\Lambda(v)$ \citep{HIRT98}. Hereafter $p(v)=L'(v)$ will stand for the \emph{prelogarithmic} impulsion function.

We emphasize that, although the calculations presented below in the rest of Section~\ref{sec:straightdef} are expressed in scalar form under the simplifying assumption $\bbeta = \eta \bhb$, this choice is made solely for notational convenience and to facilitate comparisons with earlier literature on dislocations. The assumption is not essential: the fundamental objects of the theory are the traction vector $\bt$ and the matrices $\sfL$ and $\sfK$, rather than the resolved stress $\sigma_{\rm r}$ and the scalars $L(v)$ and $\kappa$ below. For accurate applications, the full vectorial character of $\bbeta$ must be retained \citep{HUNT11}.

%\MJ{As a side remark, we underline that computations in Sections \ref{sec:steady}, \ref{sec:isotrop}, and \ref{sec:dsrstf} are actually valid without the contractions with $\bb$ on the level of $\sfL$, $\bt$ and $\bbeta$.\footnote{\MJ{On the one hand, notice that $\bbeta=\bhb\,\eta$ is actually an \textit{assumption} on the level of the dislocation shape, which infers its local behaviour from its global one, and might not be satisfied in general. On the other hand, $\sigma_{\rm r}:=\bt\cdot\bhb$ can be rather seen as a projection of the traction in a particular direction $\bhb$.}}
%This involves $\sfL(v)$ instead of $1/b^2 L(v)$ and avoids the projections $\sigma_{\rm r}:=\bt\cdot\bhb$ and $\bbeta=\bhb\,\eta$ in the subsequent computations. In such case, these provide the steady state, isotropic-like form, and dynamical stress response in space-time form for general straight dislocations in anisotropic media. In particular, \eqref{eq:sskern}, \eqref{eq:sigwithAB}, and \eqref{eq:sigisolike} below can also be recast in a more general matrix form.}

%******************************************************************************************
\subsection{Steady state}
\label{sec:steady}
An expression of the stress kernel of the Weertman equation in terms of $L(v)$ is derived hereafter, thereby proving in anisotropic elasticity a connection between the analytic continuation of $L(v)$ to the upper complex $v$ plane, and the velocity-dependent coefficients of the Weertman equation. 

We proceed from the scalar expression \eqref{eq:sigr_scal}. For uniform source motion at velocity $v$ along direction $\bbm$, 
\begin{align}
\eta(k_x,\omega)=(2\pi)\eta(k_x)\delta(\omega-k_x v),
\end{align}
where $\eta(k_s)$ defines the steady master shape of the defect; see, e.g., \cite{PELL18} for explicit examples. The frequency inversion of \eqref{eq:sigr_scal} can then be done right away. The calculation is reduced to the co-moving frame of the dislocation by omitting the trivial uniform-translation factor $\ee^{-\ii k_x v t }$, yielding the stress as
\begin{align}
\label{eq:tractvseta1dvconst}	
\sigma_{\rm r}(k_x)
&=\frac{2\pi}{b^2} |k_x|\,L\left(\sign(k_x)v+\ii 0^+\right)\eta(k_x)\hspace{4cm}(1D).	
\end{align}
Its Fourier inversion follows from separating positive and negative wave modes, writing 
\begin{align}
\sigma_{\rm r}(k_x)
&=\frac{2\pi}{b^2} \left[k_x\,L\left(v+\ii 0^+\right)\theta(k_x)-k_x\,L\left(v-\ii 0^+\right)\theta(-k_x)\right]\eta(k_x),	
\end{align}
and invoking identities \eqref{eq:heisen}. One readily deduces the following general form of the steady-state kernel:
\begin{align}
\label{eq:sskern}
\sigma_{\rm r}(x)&=\frac{2}{b^2}\Re\int_{-\infty}^{+\infty}\dd x'\frac{L(v+\ii 0^+)}{x-x'+\ii 0^+}\frac{\partial\eta}{\partial x}(x'),
\end{align}
which has not previously appeared in the literature.

%******************************************************************************************
\subsection{Isotropic-like form}\label{sec:isotrop}
In the context of the Peierls-Nabarro model, it is convenient to recast Eq.\ \eqref{eq:sskern} in an isotropic-like form. This allows for a direct transfer of previously established results---such as the mobility law or the equation of motion of the dislocation---to the anisotropic case. To this end, we introduce a reference shear modulus $\mu$; e.g., the relaxed or unrelaxed in-plane shear modulus \citep{ROUN99}. A reference shear wave speed $c$, and a reference line energy density $w_0$ are then defined as
\begin{align}
\label{eq:mucw0def}
c:=\sqrt{\mu/\rho},\qquad w_0:=\mu b^2/(4\pi).
\end{align}

Writing \eqref{eq:sskern} as
\begin{align}
\label{eq:sigwithL}
\sigma_{\rm r}(x)&=\frac{\mu}{\pi}\frac{1}{2 w_0}\Re\int_{-\infty}^{+\infty}\dd x'\frac{L(v+\ii 0^+)}{x-x'+\ii 0^+}\frac{\partial\eta}{\partial x}(x'),
\end{align}	
introducing as in the isotropic case dimensionless coefficients $A(v)$ and $B(v)$ defined by  %\citep{PELL12}
\begin{align}\label{def:A-B}
\frac{L(v+\ii 0^+)}{2 w_0}&:=-A(v)+\ii\, B(v),	
\end{align}
and invoking the Plemelj identity \eqref{eq:plemelj}, transforms equation \eqref{eq:sigwithL} into
\begin{align}
\label{eq:sigwithAB}
\sigma_{\rm r}(x)&=-\frac{\mu}{\pi}A(v)\pv\int_{-\infty}^{+\infty}\frac{\dd x'}{x-x'}\frac{\partial\eta}{\partial x}(x')+\mu B(v)\frac{\partial\eta}{\partial x}(x),
\end{align}	
where `$\pv$' is the principal value. This expression recovers the stress kernel in the Weertman equation \citep{ROSA01}, 
but now with generic anisotropic definitions for $A(v)$ and $B(v)$.

The two infinitesimals in \eqref{eq:sigwithL} have distinct physical interpretations. As previously discussed, the $\ii 0^+$ in $L(v + \ii 0^+)$ arises from the PLA and enforces causality, thereby accounting for radiative effects in intersonic and supersonic regimes via the coefficient $B(v)$. In contrast, the $\ii 0^+$ in the denominator serves a different purpose. The kernel describes the response of a Volterra dislocation (the response of the full dislocation being obtained through convolution), and for such a dislocation to emit radiation it must possess a vanishingly small but finite core size, $a \to 0^+$, rather than being strictly coreless. Accordingly, this $\ii 0^+$ should be interpreted as $\lim_{a \to 0^+} (\ii a)$.
%******************************************************************************************
\subsection{Dynamical stress response in space-time form}
\label{sec:dsrstf}
Next, the space-time representation of the anisotropic stress-response kernel is derived and cast in isotropic-like form. This involves elucidating the large-velocity behavior. For isotropic elasticity, Eqs.\ \eqref{eq:liminf} (in a different but equivalent form), \eqref{eq:sigisolike}, and \eqref{eq:kexpr0} below  have been found crucial to the scalar equation of motion of dislocations \citep{PELL12,PELL14}. We rewrite Eq.\ \eqref{eq:sigr_scal} as
\begin{align}
\label{eq:sigr_scal2}
\sigma_{\rm r}(k_x,\omega)&=\frac{\mu}{2 w_0}|k_x|L\left(\frac{\omega}{|k_x|}+\ii 0^+\right)\eta(k_x,\omega):=\left[-\frac{\mu}{\pi}K(k_x,\omega)(\ii \,k_x)+\ii\omega\frac{\mu}{2 c}\kappa\right]
\eta(k_x,\omega),
\end{align}
where we have introduced the scalar dimensionless radiative coefficient
\begin{align}
\label{eq:kappadef}
\kappa&:= \frac{2 c}{\mu}\,\bhb\cdot\sfK\cdot{\bhb},
\end{align}
from  $\sfK$ in \eqref{eq:sfKdef}, and the in-plane response kernel defined as
\begin{align}
\label{eq:inplane_def}
K(k_x,\omega)&:=\frac{\ii\pi}{2 w_0}\sign(k_x)\left[L\left(\frac{\omega}{|k_x|}+\ii 0^+\right)-\ii w_0 \frac{\kappa}{c}\frac{\omega}{|k_x|}\right].
\end{align}
Moreover, combining Eqs.~\eqref{eq:limitkzer}, \eqref{eq:scalLdef}, and \eqref{eq:kappadef} one finds that $L(v)$ behaves in the limit $|v|\to\infty$ as
\begin{align}
\label{eq:liminf}
L(v)=\ii w_0\frac{\kappa}{c}v\sign(\Im v)+\mathop{\text{o}}(1).	
\end{align}
Here, the decay of the non-constant next-to-leading-order term ensures that in the Fourier inversion of $K(k_x,\omega)$ below, boundary contributions vanish in partial integrations, and that Jordan’s lemma holds in Cauchy integrals. The detailed proof of this decay is provided in \ref{sec:limitLcorr}.

An inverse Fourier transform gives the space-time representation of the stress kernel \eqref{eq:sigr_scal2} in the same form as in the isotropic case  
%\citep{PELL14,PELL23}
\begin{align}
\label{eq:sigisolike}
\sigma_{\rm r}(x,t)&=-\frac{\mu}{\pi}\int_{-\infty}^{+\infty}\dd x'\int_{-\infty}^t \dd t'\,K(x-x',t-t')\frac{\partial \eta}{\partial x}(x',t')-\kappa\frac{\mu}{2 c}\frac{\partial\eta}{\partial t}(x,t),
\end{align}
where the products $\mu K(x,t)$ and $\kappa \mu/(2c)$ do not depend on our choice for $\mu$. A summary of known results for $K(x,t)$ in isotropic elasticity, covering the three dislocation characters (glide and climb edges, and screw---respectively corresponding to fracture modes II, I, and III) is provided in Appendix A of \cite{PELL23}. 

Explicit isotropic calculations support the following functional relation between kernel $K(x,t)$ and the impulsion $p(v)=L'(v)$:
%\citep{PELL12} 
\begin{align}
\label{eq:kexpr0}
K(x,t)&\equiv\frac{\theta(t)}{2w_0}\frac{1}{t^2}\Re p\left(\frac{x}{t}+\ii 0^+\right)
\end{align}
where $\theta(t)$ is the Heaviside unit-step function. The right-hand side defines a real-valued generalized function (GF) (or distribution) by analytic continuation of $p(v)$ to complex arguments \citetext{\citealp{BREM61},\citealp[p.\ 93]{GELF64}};\footnote{It is undefined at $t=0$. If needed, an explicit regularization such as in \citep{PELL14} could be employed.} see \ref{app:continuation} for an example.
%\footnote{\MJ{Je serais pour mettre la preuve qui suit en Annexe. Ce sont des simples manipulations, qui n'ont rien d'éclairant, pour ton propos, non?}}

We now demonstrate Eq.~\eqref{eq:kexpr0} via Fourier inversion of Eq.~\eqref{eq:inplane_def}. Starting from
\begin{align}
\mathcal{F}^{-1}[K(k_x,\omega](x,t)&=\frac{\ii\pi}{2w_0}\int\frac{\dd k_x}{2\pi}\frac{\dd\omega}{2\pi}\sign(k_x)\left[L\left(\frac{\omega}{|k_x|}+\ii 0^+\right)-\ii w_0\frac{\omega}{|k_x|}\frac{\kappa}{c}\right]\ee^{\ii k_x x -\ii\omega t},
\end{align}
we first invert with respect to $\omega$. By the PLA, the integrand has no singularities in the upper complex $\omega$-plane and vanishes at infinity, yielding a causal function that vanishes for $t<0$. We therefore consider $t>0$. Introducing $v=\omega/|k_x|$ and integrating by parts over $v$ gives
\begin{align}
\mathcal{F}^{-1}[K(k_x,\omega](x,t)
&=\frac{\ii\pi}{2w_0}\int\frac{\dd k_x}{2\pi}k_x\ee^{\ii k_x x}\int\frac{\dd v}{2\pi}\left[L\left(v+\ii 0^+\right)-\ii w_0\frac{\kappa}{c}v\right]\ee^{-\ii|k_x|v t}\\
&=-\frac{\pi}{2w_0 t}\int\frac{\dd k_x}{2\pi}\sign(k_x)\ee^{\ii k_x x}
\biggl(
\frac{1}{2\pi}\left\{
\left[L\left(v+\ii 0^+\right)-\ii w_0\frac{\kappa}{c}v\right]\ee^{-\ii|k_x|v t}
\right\}_{v=-\infty}^{v=+\infty}\nonumber\\
&\hspace{1cm}\biggl.{}-\int\frac{\dd v}{2\pi}\left[p\left(v+\ii 0^+\right)-\ii w_0\frac{\kappa}{c}\right]\ee^{-\ii|k_x|v t}\biggr),
\end{align}
where the boundary terms vanish due to Eq.~\eqref{eq:liminf}. Exchanging the order of integration and expressing the sign function in terms of Heaviside functions yields
\begin{align}
&\mathcal{F}^{-1}[K(k_x,\omega](x,t)
=\frac{\pi}{2w_0 t}\int\frac{\dd v}{2\pi}\left[p\left(v+\ii 0^+\right)-\ii w_0\frac{\kappa}{c}\right]\int\frac{\dd k_x}{2\pi}\sign(k_x)\ee^{\ii k_x x-\ii|k_x|v t}\nonumber\\
&=\frac{\pi}{2w_0 t}\int\frac{\dd v}{2\pi}\left[p\left(v+\ii 0^+\right)-\ii w_0\frac{\kappa}{c}\right]
\left[\int\frac{\dd k_x}{2\pi}\theta(k_x)\ee^{\ii k_x(x-vt)}-\int\frac{\dd k_x}{2\pi}\theta(-k_x)\ee^{\ii k_x(x+vt)}\right].
\end{align}
Using the Fourier transforms of Heaviside functions (Eq.~\eqref{eq:heisen}), this reduces to
\begin{align}
\mathcal{F}^{-1}[K(k_x,\omega](x,t)
&=\frac{1}{4w_0 t^2}\int\frac{\dd v}{2\ii\pi}\left[p\left(v+\ii 0^+\right)-\ii w_0\frac{\kappa}{c}\right]
\left[\frac{1}{v-(x/t+\ii 0^+)}-\frac{1}{v-(-x/t+\ii 0^+)}\right].
\end{align}
Finally, evaluating the Cauchy integral with a counter-clockwise contour closed in the upper complex $v$-plane and applying Jordan’s lemma gives
\begin{align}
\mathcal{F}^{-1}[K(k_x,\omega](x,t)
&=\frac{1}{4w_0 t^2}\left[p(x/t+\ii 0^+)-p(-x/t+\ii 0^+)\right]=\frac{1}{2 w_0}\frac{1}{t^2}\Re p(x/t+\ii 0^+),
\end{align}
where we used the odd property $p(-v) = -p(v)$. This establishes Eq.~\eqref{eq:kexpr0} for $t>0$ $\square$.

%%%%%%%%%%%%%%%%%%%%%%%%%%%%%%%%%%%%%%%%%%%%%%%%%%%%%%%%%%%%%%%%%%%
\section{Prelogarithmic Lagrangian factor and radiation}
\label{sec:plfr}
The fundamental definition of the prelogarithmic Lagrangian factor $L(v)$ is rooted in energy considerations \citep{BELT68,HIRT98}. This section aims to clarify the interpretation of the imaginary part in $L(v+\ii 0^+)$, by articulating the dual nature of $L(v)$ being related both to energies and, as in the foregoing, to the stress response function. Our analysis starts from the most fundamental level of the Green's function \citep{MURA87,BUDR93,KOSL02,PELL17a}. In this Section, $\bx:=(x,y,z)$, and $\bk$ stands for a three-dimensional Fourier vector, unless otherwise stated.

\subsection{Eigendecompostion of the Green's function and supersonic or intersonic radiation}
In Fourier form, the Green's functions of the Navier equation are built on the following `template', where $N_{ij}(\bk)\defi k_k c_{iklj}k_l$ is the acoustic tensor:
\begin{eqnarray}
\label{eq:fgdef}
\mathsf{G}(\bk,\omega)=[\sfN(\bk)-\rho \omega^2\sfI\,]^{-1}.
\end{eqnarray}
The \emph{retarded} (denoted by $+$) and \emph{advanced} (denoted by $-$) Green's functions are defined as
\begin{eqnarray}
\label{eq:gdef}
\sfG^{\pm}(\bk,\omega)\defi\lim_{\epsilon\to 0^+}\sfG(\bk,\omega\pm\ii\epsilon)=\sfG(\bk,\omega\pm\ii 0^+),
\end{eqnarray}
with only the causal $G^+$ being physically relevant in the present context, although $G^-$ appears in the calculation of energies. Causality is implemented via the PLA; see Section \ref{sec:2dinplane}.

The spectral (eigendecomposition) form of $\sfG^\pm$ is given by
\begin{align}
\label{eq:gmodes}	
\mathsf{G}^{\pm}(\bk,\omega)&=\frac{1}{\rho}\sum_{\gamma=1}^3 \frac{\sfP^\gamma(\bhk)}{c_\gamma^2(\bhk) k^2-(\omega\pm\ii 0^+)^2},
\end{align}
where the projectors $\sfP^\gamma(\bhk):=\bhe^\gamma(\bhk)\otimes\bhe^\gamma(\bhk)$ are constructed from the eigenvectors $\bhe^\gamma(\bhk)$ of the operator 
\begin{align}
\sfC(\bhk)&:=\sfN(\bk)/(\rho k^2)=\sum_{\gamma=1}^3 c_\gamma^2(\bhk)\,\sfP^\gamma(\bhk),    
\end{align}
and the $c_\gamma(\bhk)$ are the direction-dependent wave speeds in the medium. Since $(\omega\pm \ii 0^+)^2$ is equivalent to $\omega^2\pm\ii \sign(\omega) 0^+$, using the Plemelj identity \eqref{eq:plemelj} gives
\begin{align}
\label{eq:gfourdistr}
\mathsf{G}^{\pm}(\bk,\omega)
=\frac{1}{\rho}\sum_{\gamma=1}^3\sfP^\gamma\left[\pv\frac{1}{c_\gamma^2 k^2-\omega^2}
\pm \ii\pi\sign\omega\,\delta\left(c_\gamma^2 k^2-\omega^2\right)\right],
\end{align}
For $\omega$ real, the real and imaginary parts, namely, 
\begin{align}
\label{eq:gstatdef}
G^0:=(G^++G^-)/2=\frac{1}{\rho}\pv\sum_{\gamma=1}^3\frac{\sfP^\gamma}{c_\gamma^2 k^2-\omega^2},    
\end{align}
and $\Im G^\pm$, represent the reactive and radiative components of the field, respectively---a distinction borrowed from antenna theory \citep{JACO89,PELL18}. The reactive part $G^0$ describes the non-radiative field with energy that remains locally bound to the source and moves with it, while the radiative part accounts for the energy emitted and propagated to infinity. 

For a source in uniform motion, the radiative contribution is nonzero only in intersonic or supersonic regimes of motion. Indeed, let us consider motion along the $Ox$ axis with speed $v$. This motion imposes the relation $\omega=k_x v$. In 2D with $k_y=0$ for a straight source, the wavenumber satisfies $k^2=k_x^2+k_z^2$, where $k_z$ is the component normal to the plane. Then, for mode $\gamma$, we have:
\begin{align}
\delta(c_\gamma^2 k^2-\omega^2) &=\frac{1}{c_\gamma^2}\delta\left(k_x^2+k_z^2-(v^2/c_\gamma^2)k_x^2\right)=\frac{1}{c_\gamma^2}\delta\left(k_z^2+k_x^2\left(1 - v^2/c_\gamma^2\right)\right).
\end{align}
This Dirac delta can contribute to a $\bk$-space integral only if there exists a real $k_z$ for which its argument vanishes. This is possible only when $|v|>c_\gamma$, i.e., in the supersonic regime relative to $c_\gamma$.

%%%%%%%%%%%%%%%%%%%%%%%%%%%%%%%%%%%%%%%%%%%%%%%%%%%%%%%%%
\subsection{Response functions of elastodynamic fields (uniformly moving sources)}
In linear elasticity under small deformations, plasticity can be formulated in terms of the distortion tensor $u_{i,j}$.
%, whose symmetric and antisymmetric parts represent strain and rotation, respectively. 
The plastic distortion $\beta^{\text{P}}_{ij}(\bx,t)$ (the \emph{eigenstain}) acts as the source of the elastic field. We introduce response functions linking the displacement $\bu$, elastic distortion $\smash{\beta_{ij}:=u_{i,j}-\beta^{\rm P}_{ij}}$ and stress $\smash{\sigma_{ij}:=c_{ijkl}\beta_{ij}}$ to dislocation sources. These kernels will be used to compute energy expressions \citep{KOSL02}.

A uniformly moving source such that $\beta^{\rm P}_{ij}(\bx,t)\equiv \beta^{\rm P}_{ij}(\bx-\bv t)$ translates in the Fourier domain as $\beta^{\rm P}_{ij}(\bk,\omega):=(2\pi)\delta(\omega-\bk\cdot\bv)\beta^{\text{P}}(\bk)$. Letting $\bx':=\bx-\bv t$, the retarded solution is expressed as the Fourier integral:
\begin{align}
\label{eq:uk}	
u_i(\bx,t)&=-\ii\int\frac{\dd^3 k}{(2\pi)^3}G^+_{ij}(\bk,\bk\cdot\bv)c_{jklm}k_k \beta^{\rm P}_{lm}(\bk)\ee^{\ii\bk\cdot\bx'}.
\end{align}

The elastic distortion can be written in terms of the retarded distortion response function $\mathbb{B}^+(\bk,\omega)$, whose components are defined by
\begin{align}
\label{eq:bline1}
B^+_{ijkl}(\bk,\omega)&\defi k_i G^+_{jp}(\bk,\omega) k_q c_{pqkl}-\delta_{ik}\delta_{jl},
\end{align}
such that
\begin{align}
\beta_{ij}(\bx,t)&=\int\frac{\dd^3\!k}{(2\pi)^3} B^+_{ijkl}(\bk,\bk\cdot\bv)\beta^{\rm p}_{kl}(\bk)\ee^{\ii \bk\cdot\bx'}.
\end{align}
Similarly, the stress field generated by $\beta^{\rm p}_{ij}$ is given by
\begin{align}
\label{eq:sigS}
\sigma_{ij}(\bx,t)=\int\frac{\dd^3\!k}{(2\pi)^3} S^+_{ijkl}(\bk,\bk\cdot\bv)\beta^{\rm p}_{kl}(\bk)\ee^{\ii \bk\cdot\bx'},
\end{align}
where $\mathbb{S}^+$ is the retarded stress response function, whose components are 
\begin{align}
\label{eq:splus}	
S^+_{ijkl}(\bk,\omega):= c_{ijmn}B^+_{mnkl}(\bk,\omega)=c_{ijmn}k_m  G^+_{no}(\bk,\omega)k_pc_{opkl}-c_{ijkl},
\end{align}
Replacing the superscript $+$ with $-$ or $0$ defines the corresponding advanced or stationary response functions.
 
%%%%%%%%%%%%%%%%%%%%%%%%%%%%%%%%%%%%%%%%%%%%%%%%%%%%%%%%%
\subsection{Integral form of kinetic and elastic energy densities}
Using Parseval’s identity and Eq.\ \eqref{eq:uk}, the line energy densities at constant velocity---kinetic $\mathcal{W}_{\rm K}$ and elastic $\mathcal{W}_{\rm S}$---can be formally expressed in terms of Green's functions as integrals over Fourier modes:
\begin{subequations}
\begin{align}
\label{eq:wkint}	
\mathcal{W}_{\rm K}(\bv)&\defi\frac{\rho}{2}\int \dd^3\!x\,|\partial_t \mathbf{u}(\bx,t)|^2
=\frac{1}{2}\int \frac{\dd^3\!k}{(2\pi)^3}\beta^{\rm p}_{ij}(\bk)W^{\rm K}_{ijkl}(\bk,\bk\cdot\bv)\beta^{\rm p}_{kl}(-\bk),\\
\label{eq:wsint}	
\mathcal{W}_{\rm S}(\bv)&\defi\frac{1}{2}\int \dd^3\!x\,\beta_{ij}(\bx,t)c_{ijkl}\beta_{kl}(\bx,t)
=\frac{1}{2}\int\frac{\dd^3\!k}{(2\pi)^3}\beta^{\rm p}_{ij}(\bk)W^{\rm S}_{ijkl}(\bk,\bk\cdot\bv)\beta^{\rm p}_{kl}(-\bk),
\end{align}
\end{subequations}
with the tensor kernels $\mathbb{W}^{\rm K}$ and $\mathbb{W}^{\rm S}$, of components:
\begin{subequations}
\begin{align}
W^{\rm K}_{ijkl}(\bk,\omega)&\defi\rho\,\omega^2 c_{ijmn}k_m G^+_{nq}(\bk,\omega)G^-_{qo}(\bk,\omega)k_pc_{opkl},\\
W^{\rm S}_{ijkl}(\bk,\omega)&\defi c_{mnop}B^+_{mnij}(\bk,\omega)B^-_{opkl}(\bk,\omega).
\end{align}
\end{subequations}
These expressions highlight the need to distinguish between retarded and advanced Green's functions when computing energies \citep{SHEN06}.

The integrals \eqref{eq:wkint}–\eqref{eq:wsint} are usually improper, diverging at $k=0$, or $k\to\infty$ (for Volterra dislocations), but they can be regularized by restricting the Fourier integration domain. More critically, the operators they involve are mathematically ill-defined: in particular, the radiative part of the product $G^+\,G^-=G^-\,G^+$ in $\mathbb{W}^{\rm K}$ lacks distributional meaning. Indeed, from the spectral representation \eqref{eq:gmodes}, this product becomes:
\begin{align}
\label{eq:gpgm}
\sfG^+(\bk,\omega)\cdot\sfG^-(\bk,\omega)
&:=\lim_{\epsilon\to 0}\sfG(\bk,\omega+\ii\epsilon)\cdot\sfG(\bk,\omega-\ii\epsilon)\nonumber\\
&{}=\lim_{\epsilon\to 0}\frac{1}{\rho^2}\sum_{\alpha=1}^3\frac{\sfP^\alpha}{\bigl[c_\alpha^2 k^2-(\omega+\ii\epsilon)^2\bigr]\bigl[c_\alpha^2 k^2-(\omega-\ii\epsilon)^2\bigr]}.
\end{align}
As $\epsilon\to 0$, using the delta-sequence representation $\pi^{-1}\epsilon/(x^2+\epsilon^2)\to \delta(x)$ \citep{KANW04}, the denominator yields:
\begin{align}
&\frac{1}{\bigl[c_\alpha^2 k^2-(\omega+\ii\epsilon)^2\bigr]\bigl[c_\alpha^2 k^2-(\omega-\ii\epsilon)^2\bigr]}
=\frac{1}{4 c_\alpha k\, \omega}\left[\frac{1}{(c_\alpha k -\omega)^2+\epsilon^2}-\frac{1}{(c_\alpha k+\omega)^2+\epsilon^2}\right]\nonumber\\
&\mathop{\simeq}_{\epsilon\to 0}\frac{\pi}{4 c_\alpha k \omega}\left[\delta(\omega-c_\alpha k)-\delta(\omega+c_\alpha k)\right]\frac{1}{\epsilon}
+\Pf\frac{1}{(c_\alpha^2 k^2-\omega^2)^2}+\bigO{\epsilon}\nonumber\\
\label{eq:radiaGG}
&=\frac{\pi}{2|\omega|}\delta\bigl(c_\alpha^2 k^2-\omega^2\bigr)\frac{1}{\epsilon}+\Pf\frac{1}{(c_\alpha^2 k^2-\omega^2)^2}+\bigO{\epsilon},
\end{align}
where $\Pf$ denotes Hadamard's finite part. The $1/\epsilon$-divergent term is undefined as a distribution, and thus so is $W_{ijkl}^K$ in this form. The problem arises from mixing radiative components of the Green's functions in the energy calculation. 

A problem of the sort is expected: energy is a real-valued scalar and cannot simultaneously represent both the localized, reactive, and conserved, field bound to the source, and the losses via the radiated field escaping to infinity---they are physically distinct in nature. Consequently, the kinetic and elastic energies at constant velocity are strictly meaningful only in the subsonic regime, where no radiation occurs and the energies are real. Nevertheless, these quantities can be formally extended to intersonic or supersonic regimes, provided the Green's functions $G^\pm$ are replaced by their shared reactive part $G^0$, thus solely retaining reactive contributions and eliminating the problematic radiative ones.

\subsection{Lagrangian and stress response function}
\label{sec:lagsection}
In contrast, the Lagrangian, defined as the difference between kinetic and elastic energy densities,
\begin{align}
\label{eq:lag}
\mathcal{L}(\bv)&:=\mathcal{W}_{\rm K}(\bv)-\mathcal{W}_{\rm S}(\bv)=\frac{1}{2}\int \frac{\dd^3\!k}{(2\pi)^3}\beta^{\rm p}_{ij}(\bk)L_{ijkl}(\bk,\bk\cdot\bv)\beta^{\rm p}_{kl}(-\bk),
\end{align}
with associated kernel
\begin{align}
\label{eq:L0def}	
L_{ijkl}&:=W^{\rm K}_{ijkl}-W^{\rm S}_{ijkl}
\end{align}
does not suffer from the pathological behavior described above. The radiative singularities present in $W^{\rm K}_{ijkl}$ and $W^{\rm S}_{ijkl}$ cancel out in the difference, leaving a well-defined expression. It follows, incidentally, that $W^S_{ijkl}$  exhibits the same singular behavior as $W_{ijkl}^K$. 

We show this and demonstrate that \emph{the Lagrangian kernel \eqref{eq:L0def} can be identified with the reactive part of the stress response function}, using the following exact manipulations, performed with finite $\epsilon$. Start with:
\begin{align}
W^{\rm S}_{mnop}&
=c_{ijkl} B^+_{ijmn}B^-_{klop}=(G^+_{iq}k_j k_r c_{qrmn}-\delta_{im}\delta_{jn})c_{ijkl}(G^-_{ks}k_l k_t c_{stop}-\delta_{ko}\delta_{lp})\nonumber\\
&=c_{mnrq}k_r(G^+_{qi}N_{ik}G^-_{ks}-2 G^0_{qs})k_t c_{stop}+c_{mnop},
\end{align}
where $N_{ij}$ has been defined in connection with \eqref{eq:fgdef}, and where  $G^0$ is defined in \eqref{eq:gstatdef}.
Next, expanding the operator product yields
\begin{align}
W^{\rm S}_{mnop}&=c_{mnrq}k_r\Bigl\{
\frac{1}{2}G^+_{qi}[N_{ik}-\rho(\omega+\ii\epsilon)^2\delta_{ik}]G^-_{ks}+\frac{1}{2}G^+_{qi}[N_{ik}-\rho(\omega-\ii\epsilon)^2\delta_{ik}]G^-_{ks}\nonumber\\
&\hspace{6.5cm}+\rho(\omega^2-\epsilon^2) G^+_{qi}G^-_{is}-2 G^0_{qs}\Bigr\}k_tc_{stop}+c_{mnop}\nonumber\\
&=c_{mnrq}k_r\Bigl\{
\frac{1}{2}G^+_{qi}(G^+)^{-1}_{ik}G^-_{ks}+\frac{1}{2}G^+_{qi}(G^-)^{-1}_{ik}G^-_{ks}
+\rho(\omega^2-\epsilon^2) G^+_{qi}G^-_{is}-2 G^0_{qs}\Bigr\}k_tc_{stop}+c_{mnop}\nonumber\\
&=c_{mnrq}k_r\left[\rho(\omega^2-\epsilon^2) G^+_{qi}G^-_{is}-G^0_{qs}\right] k_tc_{stop}+c_{mnop}\nonumber\\
\label{eq:bsred}
&=W^{\rm K}_{mnop}-(c_{mnrq}k_r G^0_{qs}k_tc_{stop}-c_{mnop})-(\epsilon^2/\omega^2) W^{\rm K}_{mnop}.
\end{align}
Taking the limit $\epsilon\to 0$ and using definition \eqref{eq:L0def}, one gets:
\begin{equation}
\label{eq:noyauL}
L_{ijkl}(\bk,\omega) = c_{ijmn}k_m G^0_{no}(\bk,\omega)k_pc_{opkl}-c_{ijkl}.
\end{equation}
Using the above expression allows us to extend $L(\bk,\omega)$ to complex values $\omega$, using \eqref{eq:gmodes} as a definition for $G^0$ for $\Im(\omega) \neq 0$. This non-singular expression, which is linear in the Green function (unlike the energy terms), coincides with the reactive part of the stress response kernel $S^+_{ijkl}(\bk,\omega)$ for $\Im(\omega) > 0$, as seen by comparing with \eqref{eq:splus}. For real $\bk$ and $\omega$, it is now clear that, as claimed,
\begin{align}\label{eq:78}
L_{ijkl}(\bk,\omega)=\Re S^\pm_{ijkl}(\bk,\omega).
\end{align}
Moreover, using \eqref{eq:noyauL} as a defining equation for $L_{ijkl}(\bk,\omega)$ for $\omega \in \mathbb{C}$, the stress operator can be obtained as
\begin{align}\label{eq:79}
S^+_{ijkl}(\bk,\omega)\equiv L_{ijkl}(\bk,\omega+\ii 0^+).
\end{align}
Therefore, the Lagrangian kernel, supplemented by the PLA, provides the retarded stress-response kernel. Accordingly, the stress field in the medium, Eq.\ \eqref{eq:sigS}, can also be expressed as
\begin{align}
\label{eq:sigL}
\sigma_{ij}(\bx,t)=\int\frac{\dd\omega}{2\pi}\int\frac{\dd^3\!k}{(2\pi)^3} L_{ijkl}(\bk,\omega+\ii 0^+)\beta^{\rm p}_{kl}(\bk,\omega)\ee^{\ii( \bk\cdot\bx-\omega t)}.
\end{align}

We close this Section by an interpretation of the Stroh formalism for plane sources of the form
\begin{align}
\beta_{ij}^{\rm P}(\bx,t)&=n_i \eta_j(\br,t)\delta(z),
\end{align}
with $\br\defi(x,y)$. Restricting to the plane $z=0$, and letting now $\bk:=(k_x,k_y)$, the traction $\bt=\bn\cdot\bsigma$ deduced from Eq.\ \eqref{eq:sigL} has Fourier components
\begin{align}
t_i(\bk,\omega)&=\left[\int\frac{\dd k_z}{2\pi}n_j L_{ijkl}(\bk,k_z,\omega+\ii 0^+)n_k\right]\eta_l(\bk,\omega)
\end{align}
Definition  \eqref{eq:Ldef} of the kernel $\mathsf{L}(\bhk,v)$ makes it clear that
the result must match \eqref{eq:tractvseta2}, implying the identity
\begin{align}
\label{eq:identityL1}
L_{il}(\bhk,\omega/k)\equiv\frac{1}{(2\pi)k}\left[\int_{-\infty}^{+\infty}\frac{\dd k_z}{2\pi}n_j L_{ijkl}(\bk+k_z\bn,\omega+\ii 0^+)n_k\right].    
\end{align}
Thus, the Stroh formalism bypasses the integral over the normal component $k_z$ when computing the traction field generated by a plane source using the Green's operator.  

\subsection{Prelogarithmic Lagrangian factor}
\label{sec:prelagsection}
For completeness and to bring things full circle with Fourier-domain calculations, the prelogarithmic Lagrangian factor $L(v)$ in \eqref{eq:prelogdef} is extracted from the integration over Fourier modes in \eqref{eq:lag}, which requires considering a straight defect moving with velocity $\bv=v\, \bbm$ where the unit vector $\bbm$ is in-plane along axis $Ox$. Taking $k_y=0$, the integration over $\bk$ is then restricted to the sagittal plane $\bk=(k_x,k_z)=k(\cos\phi,\sin\phi)$ with $\phi\in[0,2\pi]$. Thus, one considers the 2D integral
\begin{align}
\label{eq:lag2d}
\mathcal{L}(\bv)&:=\frac{1}{2}\int \frac{\dd^2\!k}{(2\pi)^2}\beta^{\rm p}_{ij}(\bk)L_{ijkl}(\bk,\bk\cdot\bv)\beta^{\rm p}_{kl}(-\bk),
\end{align}
along with the plastic eigenstain of a Volterra dislocation of Burgers vector $\bb$, namely,
\begin{align}
\beta^{\rm P}_{ij}(\bk)\defi\frac{\ii\,n_i b_j}{k_x+\ii 0^+}.
\end{align}
Here $0^+$ represents here the reciprocal of a very large system size to regularize the integral as $k_x\to 0$.\footnote{However this prescription turns out unnecessary, as the function is to be multiplied by $k_x$ owing to \eqref{eq:nLn0} below.} Equivalently, one could impose a lower cut-off on $k$. 
%The calculation differs from the traditional approach in that it proceeds in the Fourier representation, without the need for knowing the solution in advance. As such it provides a basis for considering dislocations more complex than Volterra ones, such as in \citep{PELL18}. 

One furthermore introduces as usual modified elastic constants $c'_{ijkl}=c_{ijkl}-\rho\,v^2\delta_{il}m_j m_k$, and slightly change the meaning of notation  \eqref{eq:notabraces} into $(\ba\bb)_{ij}:=a_jc'_{ijkl}b_k$.\footnote{Depending on authors, the indexing convention may vary, but must be such that $(\bn\bn)$ does not depend on $v$ if $\bbm\cdot\bn=0$.} Then $\sfG^0(\bk,\bk\cdot\bv)=\pv\,(\bk\bk)^{-1}$, and using \eqref{eq:noyauL} and the orthogonality of $\bn$ and $\bbm$, the integrand in \eqref{eq:lag2d} features the tensor kernel
\begin{align}
n_jL_{ijkl}n_k&=\left[(\bn\bhk)\cdot(\bhk\bhk)^{-1}\cdot(\bhk\bn)-(\bn\bn)\right]_{il}.
\end{align}
The angular integral over $\phi$ is done by introducing the unit vector $\bl=\dd\bhk/\dd\phi$ and expanding $\bn$ over the rotating orthogonal basis $(\bhk,\bl)$. Using the fact that $\bn\cdot\bl=\bbm\cdot\bhk=k_x/k$ one gets
\begin{align}
\label{eq:nLn0}
n_jL_{ijkl}n_k
&=(\bn\cdot\bl)^2\left[(\bl\bhk)\cdot(\bhk\bhk)^{-1}\cdot(\bhk\bl)-(\bl\bl)\right]_{il}
=\frac{k_x^2}{k^2}\left[(\bl\bhk)\cdot(\bhk\bhk)^{-1}\cdot(\bhk\bl)-(\bl\bl)\right]_{il}.
\end{align}
Since $\bhk\cdot\bl=0$, identity \eqref{eq:pLLident} of the Stroh formalism applies, to express the above as
\begin{align}
\label{eq:nLn}
n_jL_{ijkl}n_k
&=-\frac{k_x^2}{k^2}\sum_{\alpha=1}^6 p_\alpha(\phi)L^\alpha_i L^\alpha_l,
\end{align}
where the vectors $\bL^\alpha$ are independent of angle $\phi$. The principal-value prescription in $\pv(\bk\bk)^{-1}$ has been omitted in the above equations. Substituting \eqref{eq:nLn} under the integral in \eqref{eq:lag2d}, and denoting the angular average by $\langle f(\phi)\rangle_\phi:= (2\pi)^{-1}\pv\int_0^{2\pi}\dd\phi f(\phi)$, immediate simplifications entail that
\begin{align}
\label{eq:lastcalc}
\mathcal{L}(v)&=\bb\cdot\left[
-\frac{1}{4\pi}\sum_{\alpha=1}^6 \langle p_\alpha(\phi)\rangle_\phi \bL^\alpha\otimes\bL^\alpha\,\right]\cdot\bb \int \frac{\dd k}{k}.
\end{align}
The well-known result $\langle p_\alpha(\phi)\rangle_\phi=\ii s_\alpha$, where $s_\alpha=\sign\Im p_\alpha$ if $\Im p_\alpha\not = 0$, and $=0$ otherwise \citep{BARN73b,TANU07}, restricts the sum in \eqref{eq:lastcalc} to the subsonic modes. The diverging integral $\int \dd k/k:=\log(R/r_0)$ is regularized by restricting $k$ to the range $\propto [1/R,1/r_0]$ (the exact prefactor is irrelevant) with $r_0\ll R$, which implements the usual inner and outer cut-off radii $r_0$ and $R$. The set of expressions \eqref{eq:prelogdef} with \eqref{eq:scalLdef} has thus been retrieved by a direct computation in the Fourier domain.  

%%%%%%%%%%%%%%%%%%%%%%%%%%%%%%%%%%%%%%%%%%%%%%%%%%%%%%%%%%%%%%%%%%%
\section{Conclusion}
\label{sec:concl}
The elastodynamic theory of traction stresses generated by systems of moving planar cracks or dislocations in anisotropic media has been revisited, and reformulated to explicitly account for radiative effects, using a Fourier-transform treatment of the in-plane position variable combined with the Stroh formalism. The analysis highlights the central role of the Lagrangian kernel and, in the general anisotropic case, confirms the expression in space–time variables previously proposed for the kernel of the Dynamic Peierls Equation. The coefficient of its instantaneous radiative term has been derived for anisotropic media and its physical origin clarified. The theory is well-suited for numerical implementations. Finally, as an immediate perspective, the formulation \eqref{eq:sigisolike} of the stress kernel of the Dynamical Peierls Equations in terms of $p(v)=L'(v)$ makes the Collective-Variable-Approximation equation of motion for a dislocation \citep{PELL14}, immediately transferable to the anisotropic case, \emph{mutatis mutandis}, using derivatives $p(v)$ and $m(v)=p'(v)$ (the mass function) computed following \cite{MALE70c}. 

\section*{CRediT authorship contribution statement}
{\bf Yves-Patrick Pellegrini:} Conceptualization, Methodology, Formal analysis, Writing -- original draft \& editing, Supervision; {\bf Marc Josien:} Conceptualization, Methodology, Formal analysis, Writing -- original draft \& editing, Supervision, Funding acquisition. {\bf Martin Chassard:} Methodology, Formal analysis, Writing -- original draft \& editing.

\section*{Declaration of competing interest}
The authors declare that they have no known competing financial interests
or personal relationships that could have influenced the work reported in this
paper.

\section*{Acknowledgements}
This work was partially supported by the CEA’s internal Programme Transverse de Comp\'etences (PTC) under the project \textbf{SiPaDD}. The work of M.C.\ was supported by the CEA Cadarache, 

%%%%%%%%%%%%%%%%%%%%%%%%%%%%%%%%%%%%%%%%%%%%%%%%%%%%%%%%%%%%%%%%%%%
\appendix
%%%%%%%%%%%%%%%%%%%%%%%%%%%%%%%%%%%%%%%%%%%%%%%%%%%%%%%%%%%%%%%%%%%
\section{Fourier transforms}
\label{sec:UFT}
Our space-time FT conventions are
\begin{align}
\label{eq:fourconv}	
f(x,t)&=\int_{-\infty}^{+\infty}\int_{-\infty}^{+\infty}\frac{\dd k}{(2\pi)}\frac{\dd \omega}{(2\pi)}f(k,\omega)\,\ee^{\ii(kx-\omega t)},\quad
f(k,\omega)=\int_{-\infty}^{+\infty}\int_{-\infty}^{+\infty}\dd x\,\dd t\,f(x,t)\,\ee^{-\ii(kx-\omega t)}.
\end{align}
where $k$ is a wavemode and $\omega$ is the angular frequency. We recall a few straightforward and useful FT's of generalized functions \citep{KANW04}. Let $\theta(x)$ denote the Heaviside (unit-step) function. One has the FTs 
\begin{align}
\label{eq:heisen}
\mathcal{F}\left[\frac{1}{\pi}\frac{1}{x+\ii 0^+}\right](k)&=-2\,\ii\,\theta(k),\qquad
\mathcal{F}\left[\frac{1}{\pi}\frac{1}{x-\ii 0^+}\right](k)= 2\,\ii\,\theta(-k).
\end{align} 
The well-known FT of the long-range kernel of the Peierls-Nabarro and Weertman equations, namely, 
\begin{align}
\label{eq:ftpv}
\mathcal{F}\left[\frac{1}{\pi}\pv\frac{1}{x}\right](k)&=-\ii\sign(k),
\end{align}	
follows from taking the mean of both Eqs.\ \eqref{eq:heisen}, and using 
$\sign(k)=\theta(k)-\theta(-k)$, and the Plemelj identity 
\begin{align}
\label{eq:plemelj}
\frac{1}{x\pm\ii 0^+}&=\pv\frac{1}{x}\mp\ii\pi\delta(x).   
\end{align}
\section{Perturbation of the Stroh eigenvalues}
\label{app:pert}
To examine the effect of the presence of the imaginary infinitesimal $\ii 0^+$ on the roots $p_\alpha$ of the sextic equation, let $\mathsf{P}\defi(\bhk\bhk)$, $\mathsf{Q}\defi(\bhk\bn)+(\bn\bhk)$, $\mathsf{R}\defi(\bn\bn)$, and 
\begin{align}
	\mathcal{A}_\alpha&\defi\mathsf{P}+\mathsf{Q}\,p_\alpha+\mathsf{R}\,p_\alpha^2-\rho v^2\,\sfI,\\
	\mathcal{A}'_\alpha&=\mathsf{Q}+2\,\mathsf{R}\,p_\alpha.
\end{align}
Then Eq.\ \eqref{eq:sextic} reads $\det\mathcal{A}_\alpha=0$.
Using Jacobi's differentiation formula $\delta\det\mathcal{A}=\mathop{\rm Tr}\left[\mathop{\rm adj}(\mathcal{A})\cdot\delta\mathcal{A}\right]$, where $\mathop{\rm adj}(\mathcal{A})$ is the adjugate of $\mathcal{A}$, the perturbation $\delta p_\alpha$  on the solution $p_\alpha$ induced by the perturbation $\delta v\defi\ii\epsilon$ on $v$ is readily found to order $\mathop{\rm O}(\epsilon)$ as
\begin{align}
\label{eq:perturb}	
\delta p_\alpha &=2\,\rho\,v\,\mathop{\rm Tr}\left[\mathop{\rm adj}(\mathcal{A}_\alpha)\right]/
\mathop{\rm Tr}\left[\mathop{\rm adj}(\mathcal{A}_\alpha)\cdot\mathcal{A}'_\alpha\right]\delta v.
\end{align}
However, the sign of this imaginary term remains unknown.

%%%%%%%%%%%%%%%%%%%%%%%%%%%%%%%%%%%%%%%%%%%%%%%%%%%%%%%%%%%%%%%\appendix
\section{Analytic continuation and generalized functions}
\label{app:continuation}
As an example, consider the screw dislocation in isotropic elasticity where, with $v\in\mathbb{C}$, \begin{align}
	\label{eq:lpscrew}	 
	L(v)&=-w_0\sqrt{1-v^2/\cS^2},\qquad p(v)=\frac{w_0}{\cS}\frac{v/\cS}{\sqrt{1-v^2/\cS^2}},	
\end{align}
and $\cS$ is the shear wave speed. The GF associated with it is,  for $v$ real,
\begin{align} 
	\label{eq:impsgf}	
	p(v)&=\frac{w_0}{\cS}\frac{v}{\cS}\left(1-v^2/\cS^2\right)^{-1/2}_+,	
\end{align}
where $(x)^\alpha_+=x^\alpha$ if $x>0$ and $0$ otherwise for $0<\alpha<1$ \citep{KANW04}. Here the function $p(v)$ of the complex variable $v$ in \eqref{eq:lpscrew}${}$ is associated with the GF \eqref{eq:impsgf}, which identifies with $\Re p(v\pm \ii 0^+)$. Indeed, if $|v|>\cS$, $\smash{p(v\pm\ii 0^+)=\pm \ii (w_0/\cS^2)|v|/\sqrt{v^2/\cS^2-1}}$, whence the result. By abuse of notation, we have used here the same symbol $p$ to denote the function and its associated GF. See, e.g., \citep{PELL15} for further examples relevant to elastodynamic Green's functions. 

%%%%%%%%%%%%%%%%%%%%%%%%%%%%%%%%%%%%%%%%%%%%%%%%%%%%%%%%%%%%%%%
\newcommand{\wcalN}{{\widetilde{\mathcal{N}}}}
\newcommand{\wbzeta}{{\widetilde{\bzeta}}}
\newcommand{\wbA}{{\widetilde{\bA}}}
\newcommand{\wbL}{{\widetilde{\bL}}}
\newcommand{\wsfL}{{\widetilde{\sfL}}}

\section[Vanishing next correction to asymptotic behavior of L(v)]{Vanishing next correction to asymptotic behavior of $L(v)$}
\label{sec:limitLcorr}
Proving the asymptotic vanishing of the next-to-leading term in \eqref{eq:liminf} amounts to showing that, with $\varepsilon:=1/v$
\begin{align}
\left. \frac{\dd\left[\varepsilon \sfL(1/\varepsilon+\ii 0^+)\right]}{\dd\varepsilon}\right|_{\varepsilon=0}=0.
\end{align}
Without loss of generality, we restrict ourselves to the case $\epsilon>0$.
Introducing rescaled vectors $\wbL^\alpha:=\varepsilon^{1/2}\bL^\alpha$ and $\wbA^\alpha:=\varepsilon^{-1/2}\bA^\alpha$, which preserves the normalization condition, the identity \eqref{eq:Ldef}	becomes
\begin{align}\label{e:D2}
\varepsilon\sfL&=\frac{1}{4\ii\pi}\sum_{\alpha=1}^6 s_\alpha \wbL^\alpha\otimes\wbL^\alpha,
\end{align}
where the signs $s_\alpha$ are constant in the limit. The $\wbA^\alpha$ and $\wbL^\alpha$ derive from the eigenvectors $\wbzeta^\alpha:=(\wbA^\alpha,\wbL^\alpha)$ of the modified matrix obtained from \eqref{def:calN}
\begin{align}
\wcalN(\varepsilon)&:=
\begin{pmatrix}
-\varepsilon (\bn\bn)^{-1}\cdot(\bn\bhk) & -(\bn\bn)^{-1} \\
-\varepsilon^2(\bhk\bn)\cdot(\bn\bn)^{-1}\cdot(\bn\bhk)+\varepsilon^2(\bhk\bhk)-\rho\sf{I} & -\varepsilon(\bhk\bn)\cdot(\bn\bn)^{-1}
\end{pmatrix},
\end{align}
whose eigenvalues $\wp_\alpha=\varepsilon p_\alpha$ are solutions of
\begin{align}
\det\left\{\varepsilon^2(\bhk\bhk)+\varepsilon[(\bhk\bn)+(\bn\bhk)]\wp_\alpha+(\bn\bn)\wp_\alpha^2-\rho\,(1+\ii 0^+)^2\sf{I}\right\}=0.
\end{align}

Differentiating \eqref{e:D2} immediately yields
\begin{align}
\label{eq:dLdvij0}
\frac{\dd}{\dd\varepsilon}[\varepsilon\sfL(\varepsilon)]&=
\frac{1}{4\ii\pi}\sum_\alpha s_\alpha
\left[
\wbL^{\alpha \prime}(\varepsilon)\otimes\wbL^\alpha(\varepsilon)
+\wbL^\alpha(\varepsilon)\otimes\wbL^{\alpha\prime}(\varepsilon)
\right].
\end{align}
Introducing the the $6\times 6$ matrix
\begin{align}
\mathcal{T}:=
\begin{pmatrix}
0&\sfI\\
\sfI&0
\end{pmatrix},
\end{align}
the derivatives at $\varepsilon=0$ read \citep{MALE70c}
%\footnote{\MJ{Suppressed the $_i$ in $\wbL^{\alpha\prime}(\varepsilon=0)$}}
\begin{align}
\wbL^{\alpha\prime}(\varepsilon=0)
&=\sum_{\beta\not=\alpha}\frac{\wbzeta^\beta\cdot\mathcal{T}\cdot\wcalN'(\varepsilon=0)\cdot\wbzeta^\alpha}{\wp_\alpha-\wp_\beta}\wbL^\beta
=\sum_{\beta\not=\alpha}\frac{\wbA^\alpha\cdot\left[\wp^\beta(\bhk\bn)+\wp^\alpha(\bn\bhk)\right]\cdot\wbA^\beta}{\wp_\alpha-\wp_\beta}\wbL^\beta,
\end{align}
where all quantities are taken at $\varepsilon=0$, and where the identity $\wbL^\alpha=-\wp_\alpha (\bn\bn)\cdot\wbA^\alpha$ has been used.

Substituting into \eqref{eq:dLdvij0}, this yields
\begin{align*}
[\varepsilon\wsfL]'(\varepsilon=0)
&=\frac{1}{4\ii\pi}\sum_{\atop{\alpha,\beta}{\alpha\not=\beta}}\frac{s_\alpha}{\wp_\alpha-\wp_\beta}\wbA^\alpha\cdot\left[\wp^\beta(\bhk\bn)+\wp^\alpha(\bn\bhk)\right]\cdot\wbA^\beta\left(\wbL^\alpha\otimes\wbL^\beta+\wbL^\beta\otimes\wbL^\alpha\right)\nonumber\\
&=\frac{1}{8\ii\pi}\sum_{\atop{\alpha,\beta}{\alpha\not=\beta}}\frac{s_\alpha-s_\beta}{\wp_\alpha-\wp_\beta}\wbA^\alpha\cdot\left[\wp^\beta(\bhk\bn)+\wp^\alpha(\bn\bhk)\right]\cdot\wbA^\beta\left(\wbL^\alpha\otimes\wbL^\beta+\wbL^\beta\otimes\wbL^\alpha\right)\nonumber\\
&=\frac{1}{4\ii\pi}\sum_{\atop{\alpha\in\calS^+}{\beta\in\calS^-}}\frac{\wbA^\alpha\cdot\left[\wp^\alpha(\bn\bhk)+\wp^\beta(\bhk\bn)\right]\cdot\wbA^\beta}{\wp_\alpha-\wp_\beta}\left(\wbL^\alpha\otimes\wbL^\beta+\wbL^\beta\otimes\wbL^\alpha\right).\nonumber
%\\
%&=\frac{1}{4\ii\pi}\sum_{\gamma_1,\gamma_2=1}^3\frac{\wbA^{\gamma_1}\cdot\left[\wp^{\gamma_1}(\bn\bhk)-\wp^{\gamma_2}(\bhk\bn)\right]\cdot\wbA^{\gamma_2+3}}{\wp_{\gamma_1}+\wp_{\gamma_2}}\left(\wbL^{\gamma_1}\otimes\wbL^{\gamma_2+3}+\wbL^{\gamma_2+3}\otimes\wbL^{\gamma_1}\right).
\end{align*}
%\footnote{\MJ{Suppressed the line with $\gamma$s which actually uses some properties written in the phrase below.}}
Using the asymptotic expressions $\wbA^\alpha=\ii\bV^\gamma/\sqrt{2\wp_\alpha\mu_\gamma}$ and $\wbL^\alpha=-\ii\sqrt{\mu_\gamma \wp_\alpha/2}\,\bV^\gamma$, for $\alpha=\gamma$ or $\alpha=\gamma+3$, and $\wp_{\gamma+3}=-\wp_\gamma$ for $\gamma=1,2,3$ (see Section \ref{sec:trltc}), 
this reduces to
\begin{align}
[\varepsilon\wsfL]'(\varepsilon=0)&=\frac{1}{4\ii\pi}\sum_{\gamma_1,\gamma_2=1}^3\frac{\bV^{\gamma_1}\cdot\left[\wp^{\gamma_1}(\bn\bhk)-\wp^{\gamma_2}(\bhk\bn)\right]\cdot\bV^{\gamma_2}}{\wp_{\gamma_1}+\wp_{\gamma_2}}\left(\bV^{\gamma_1}\otimes\bV^{\gamma_2}+\bV^{\gamma_2}\otimes\bV^{\gamma_1}\right)=0,
\end{align}
where use has been made of the antisymmetric character of the summand in the index pair $(\gamma_1,\gamma_2)$ in the last equality. Hence the result $\square$.

%%%%%%%%%%%%%%%%%%%%%%%%%%%%%%%%%%%%%%%%%%%%%%%%%%%%%%%%%%%%%%%%%%%%%%%%%%%%% REFERENCES
%%%%%%%%%%%%%%%%%%%%%%%%%%%%%%%%%%%%%%%%%%%%%%%%%%%%%%%%%%%%%%%%%%%%%%%%%%%%\

\end{document}